\documentclass[12pt,a4paper,notitlepage]{article}
\usepackage{ucs}                                                                                %
\usepackage[utf8x]{inputenc}                                                              %
\usepackage{textcomp}                                                                       %
\usepackage[english,french]{babel}                                                   %
\usepackage{amsmath,amssymb}                                                     %
\usepackage{pifont}                                                                              %
\title{Link between  New Versions of the Hierarchical Reference Theory of Liquids
and of  the Non Perturbative Renormalization Group in Statistical
Field Theory}
\author{Jean-Michel Caillol\\
LPT (UMR 8627 )\\
B\^at. 210 \\
Universit\'e de Paris-Sud \\
F-91405 Orsay CedexFrance \\
and \\
CNRS, Orsay, F-91405, France}                                                                            %
\date{\today}   
\begin{document}
\selectlanguage{english}
\maketitle
\begin{abstract}
 I propose a new version of the Hierarchical Reference Theory of liquids. Two formalisms,
one in the grand canonical ensemble, the other in the framework of statistical field  theory
are given in parallel. In the latter the theory is an avatar of a new version of the non perturbative
renormalization group (J. Phys. A : Math. Gen.  \textbf{42},  225004 (2009)). 
The flow of the Wilsonian action as well as that of the effective average action of Wetterich are derived
and a simple relation between the two functionals is established.
The  standard Hierarchical Reference Theory  for liquids
(\textit{Adv. Phys.}  \textbf{44}, 211 (1995)) is  recovered  for a sharp infra-red cut-off of the propagator 
\end{abstract}
\textbf{key-words :}statistical field theory; non-perturbative renormalization group; 
theory of liquids;  hierarchical reference theory.
\newpage
\tableofcontents
\newpage
\section{Introduction\label{intro}}
The first  theory of liquids which incorporates the ideas of the renormalization group (RG)
of Wilson \cite{Wilson,Wegner} was
proposed more than twenty-five  years ago in a series of outstanding papers by Reatto, 
Parola and co-workers who developed the so-called Hierarchical Reference Theory (HRT) 
which gives an accurate description of the thermodynamics and the structure of a wide 
class of fluids close to and away from the critical region \cite{paro1,paro2,Parola, Parola1,Reatto}.
The theory was developed and improved over years by these authors until very recently.

During roughly the same period  the Wilson approach  to the (RG) has been the subject of a revival 
in both statistical physics and field theory. Two main formulations of the non perturbative renormalization group (NPRG) 
have been developed in parallel. 
In the first one,  the RG transformations concern  the Hamiltonian  (or action) of the model
\cite{Wilson,Wegner,Polchinski,Bervillier,Morris} while, in the second,
the RG flow of the free energy is under scrutinity, or rather, to be more precise, the so-called
 ''effective average action'' \cite{Wetterich,Delamotte}.  
In both approaches, at scale ``k''  (in momentum space), a cut-off  is introduced to separate
 the slow ($q<k$) and the fast ($q>k$) modes of the field.
In the first approach the Wilsonian action is defined to be the effective action of the low energy modes yielding
 the same physics a long distances than the bare action. In this case the cut-off acts as an ultra-violet (UV) cut-off.
In the formalism of  Wetterich  these ideas are implemented at the level of the Helmholtz free energy of  the fast modes
(in fact one is rather interested  in its Legendre transform, the effective average action) and thus, in this approach,
the cut-off acts conversely  as an infra-rouge (IR) cut-off.
The two points of view can be reconciliated formally \cite{Morris,Caillol-RG}.

It turns out that many of new results obtained recently for  the NPRG had been anticipated by the seminal 
works of Reatto `et al.'.
Since the two communities have few contacts and use a different language many things have been discovered
or rediscovered independently by both sides in the ignorance of the results obtained by the other camp.
Today field theorists  would say that HRT is a sharp cut-off version 
of Wetterich effective average action approach to the RG, while specialists of liquids would claim that Wetterich theory  is nothing but
a field theoretical version of HRT with smooth cut-off and strange notations.

A precise comparison between the two worlds however requires a field theoretical representation of simple classical liquids. Technically, 
a simple Hubbard-Stratonovich transform does the job \cite{Strato,Hubb}. This has been realized by many authors
 \cite{Brilliantov,Song,KSSHE,RG-Fluids}. The more achieved scalar field theoretical representation of liquids at equilibrium
has been christened the KSSHE theory after the names of Kac \cite{Kac}, Siegert \cite{Siegert}, Stratonovich \cite{Strato},
Hubbard \cite{Hubb} and Edwards \cite{Edwards} and is developed 
 in references \cite{KSSHE,RG-Fluids} \textit{inter alia}.
I show in the latter reference  that the implementation of Wetterich effective
average action method in the framework of KSSHE theory yields indeed HRT in the ultra-sharp cut-off limit.

In general the field theory obtained via a Hubbard-Stratonovich transform yields a non-canonical theory 
in the sense that the coupling between the field $\varphi$ and the source, here the chemical potential $\nu$,
is non linear (see section~\ref{Field}
below). An additional transformation is needed to obtain a standard field theory with a linear coupling. I adhered
to this method in my previous works on liquids \cite{KSSHE,RG-Fluids}. However it turns out that the NPRG ideas 
can easily be extended to the non-canonical theory itself and it has then the advantage to yield
extremely simple functional  relationships between the Wilsonian action and Wetterich's
effective average action \cite{Caillol-RG}. In the present paper I will show
 that applying this new field theoretical formalism to the
KSSHE theory leads to a slightly different version of HRT with some advantages however.
For instance  the flow equations of Reatto `et al' are reobtained by taking the limit of an ordinary sharp cut-off.
Moreover the grand-canonical functionals corresponding to the Wilsonian
and  effective average actions of Wetterich are clearly identified which allows a deeper understanding 
of the NPRG for liquids.

The paper is organized as follows. In next section~\ref{model},  I circumscribe the type of liquids under study (the simplest!)
and give their representations in the grand-canonical ensemble and in field theory. Then, in section~\ref{CO}, the coarse-graining
operations of the various functionals of interest are detailed and their flow equations are derived in section~\ref{Flow}.
in the last paragraph~\ref{flowAA}, the usual HRT equation is obtained by taking the limit of a sharp cut-off. We summarize and conclude in 
section~\ref{conclu}. Two appendices explain some technical points.

\section{The model\label{model}}
\subsection{Grand-Canonical representation\label{GC}}
I consider a  classical  simple fluid made  of identical hard spheres (HS) of diameter $\sigma$ with, in addition,  isotropic pair interactions 
$v(r_{ij})$ ($r_{ij}=| x_i -x_j |$, $x_i$  position of particle "$i$"). Since $v(r)$ can be chosen arbitrarily  in the core,
we assume that $v(r)$ has been regularized  for $0 <r < \sigma$  in such a way that 
 its Fourier transform $\widetilde{v}_{q}$ is a well behaved function of $q$ 
and that $v(0)$ is a finite quantity. For convenience the domain $\Omega \subset \mathbb R^d$ occupied by the molecules is 
chosen to be a $d-$dimensional cube of side $L$  within periodic boundary (PB) conditions; the volume of $\Omega$ is noted $V=L^d$.
The fluid is at equilibrium in the grand canonical (GC) ensemble, $\beta=1/k_{\textrm{B}}T$ is the inverse temperature ($k_{\mathrm{B}}$
 Boltzmann's constant), and $\mu$  the chemical potential. In addition the particles are subject to an external potential $\psi(x)$ and we will denote 
by $\nu(x)=\beta (\mu-\psi(x))$ the dimensionless local chemical potential. I adhere to notations usually adopted in standard textbooks 
on liquids (see e.g. \cite{Hansen}) and  thus denote by $w(r)=-\beta v(r)$  \emph{minus } the dimensionless pair interaction. 
Moreover we  restrict ourselves to the case of \textbf{attractive} interactions, i.e. such that $\widetilde{w}(q)>0$ for all $q$.
The model admits a thermodynamic limit
if $w(r)$ decays faster than  $1/ r^{d + \epsilon}$ as $r \to  \infty $, where $d$ is the space dimensions and
 $ \epsilon > 0$ \cite{Ruelle}.

In a given GC configuration $\mathcal{C}\equiv(N;x_1 \ldots x_N)$ 
the microscopic density of particles at point $x$  reads
\begin{equation}
\label{dens}
\widehat{\rho}(x|\mathcal{C}) =
\sum_{i=1}^{N} \delta^{(d)}(x-x_i) \; ,
\end{equation}
and the grand canonical partition function (GCPF)  $\Xi\left[ \nu \right] $ which encodes 
all the physics of the model at equilibrium is defined  as \cite{Hansen}
\begin{eqnarray}
\label{csi}\Xi\left[ \nu \right] &=&
\mathrm{Tr}\left[ \; \exp\left( -\beta \mathcal{H} \left[\mathcal{C}\right] 
\right) \right] \; , \nonumber \\
-\beta \mathcal{H}&=& \frac{1}{2} 
\widehat{\rho}\cdot  w \cdot \widehat{\rho}  \, + \, 
 ( \nu- \frac{w(0)}{2} ) \,  \cdot \widehat{\rho}  \nonumber  \; ,\\ 
 \mathrm{Tr}\left[  \ldots \right] &=&
 \sum_{N=0}^{\infty}
\frac{1}{N!} \int_{\Omega}d1 \ldots dn \; \exp[-\beta V_{\mathrm{(HS)}}\left[ \mathcal{C}\right] ] \ldots \; ,
\end{eqnarray}
where $i \equiv x_i $ and $di\equiv d^{d}x_i$. 
In equation~ (\ref{csi}) $\beta V_{\mathrm{(HS)}}\left[ \mathcal{C}\right] $ denotes 
the HS contribution to the configurational integral. Its Boltzmann factor is either 0 if configuration $\mathcal{C}$ involves
overlaps of spheres, or 1 otherwise.  Note that  in fact  $\Xi\left[ \nu \right] $ does not depend 
on the self energy $ - w(0)/2$.
For a given volume $V$ and a given inverse temperature
 $\beta$, $\Xi\left[ \nu \right]$ is  a log-convex  functional of the local chemical potential $\nu(x)$ \cite{Goldenfeld,Cai-dft}.
 I have  employed here convenient  matricial notations.
\begin{eqnarray}
 \overline{\nu} \cdot \widehat{\rho} 
 &\equiv&  \int_{\Omega} d^dx  \; \overline{\nu}(x)\widehat{\rho}(x | \mathcal{C}) \\
\widehat{\rho} \cdot
w \cdot \widehat{\rho} & \equiv &
\int_{\Omega} d^dx \, d^dy \, 
\widehat{\rho}(x| \mathcal{C}) \, 
 w(x,y) \,  \widehat{\rho} \, (y|\mathcal{C}) \; .
\end{eqnarray}

\subsection{Field theoretical representation\label{Field}}
It is possible to express $\Xi [ \nu ]$ as a functional integral making thus  a link between  the theory of liquids
and statistical field theory \cite{Brilliantov,Song,KSSHE}. With the help of a Hubbard-Stratonovich 
transform \cite{Strato,Hubb} one gets:
\begin{eqnarray}
\label{attractive}
\Xi\left[ \nu \right] &=&
\mathcal{N}_{w}^{-1} \int \mathcal{D} \varphi \;
\exp \left( -\frac{1}{2}
 \varphi \cdot R \cdot  \varphi \;
 + W_{\textrm{(HS)}}\left[ \nu -w(0)/2 + \varphi\right]
 \right)
\end{eqnarray}
where $W_{\textrm{(HS)}} \equiv \ln \Xi_{\textrm{(HS)}}$,  $\varphi$ is a real random scalar field, $R \equiv w^{-1}$ 
is the inverse of $w$ (in the sense of operators, i.e. $w(1,3) \cdot R(3,2)=\delta^{(d)}(1,2)$) and
$\Xi_{\mathrm{(HS)}}\left[ \nu -w(0)/2 + \varphi\right] $ denotes the GCPF of bare hard
spheres in the presence of a local chemical potential $\nu(x) -w(0)/2  + \varphi(x)$. 
We optimistically  suppose that the functional  $\Xi_{\mathrm{(HS)}}\left[ \nu \right]$ as well as its functional derivatives,
\textit{i. e.} the density-density correlation functions,  are  known exactly. This is \textit{nearly} true since
the HS fluid has been under scrutinity since so many years \cite{Hansen}.

I have noted $\mathcal{N}_{w}$ the normalization constant
\begin{equation}
 \label{normaw}
 \mathcal{N}_{w}= \int \mathcal{D} \varphi \;  \exp \left( -\frac{1}{2}
  \varphi \cdot  R \cdot  \varphi  \right) \; .
\end{equation}
The functional integrals which enter
equations\ (\ref{attractive}) and\ (\ref{normaw}) can
be given a precise meaning in the case where the domain $\Omega$
is a cube of side $L$ with periodic boundary conditions, in this case the measure $\mathcal{D} \varphi$ 
reads \cite{Wegner}
\begin{subequations}
\label{dphi}
\begin{eqnarray}
\mathcal{D} \varphi & \equiv & \prod_{q \in \Lambda}
\frac{d \widetilde{\varphi}_{q} }
{\sqrt{2 \pi  V}} \\
d \widetilde{\varphi}_{q} d
\widetilde{\varphi}_{-q} & = & 2 \;
d\Re{\widetilde{\varphi}_{q}} \;
d\Im{\widetilde{\varphi}_{q}} \textrm{ for } q \ne
0 \; .
\end{eqnarray}
\end{subequations}
where $\Lambda = (2 \pi/L)\;  {\mathbb Z}^D$ ($ {\mathbb Z}$ ring of integers) is the reciprocal cubic
lattice. Note that the reality of $\varphi$ implies that, for $q\ne
0$, $\widetilde{\varphi}_{-q} =
\widetilde{\varphi}_{q}^{\star}$, where the star means
complex conjugation. With this slick normalization of the functional measure one has exactly
\begin{equation}
\label{NNN}
\mathcal{N}_{w}=\exp\left( \frac{V}{2} \int_{q} \ln \widetilde{w}(q)\right) 
\end{equation}
in the limit of large systems ($L\rightarrow \infty$).

The field theoretical representation~\eqref{attractive} of the GCPF $\Xi [ \nu ]$ can be extended to arbitrary pair potentials.
If $w(1,2)$ contains a definite negative part corresponding to a repulsive interaction then
a second scalar field, purely imaginary must  be introduced via an additional Hubbard-Stratonovich transform,
 see e.g. ref~\cite{KSSHE}.  Handling these two fields makes the algebra a little
bit more complicated but by no means intractable; we lazily skip this complication in this paper and restrict
ourselves to attractive $w(r)$. Alternatively, one can gather the HS and repulsive pair potentials contributions in a 
single functional $W_{\textrm{(Ref)}}$ which will play the role devoted to $W_{\textrm{(HS)}}$.

It is important to note that \eqref{attractive} ``looks'' like the generator of Green's function of a field theory with an action
$\mathcal{S}[\varphi]= \frac{1}{2}
 \varphi \cdot  w^{-1} \cdot  \varphi   - W_{\textrm{(HS)}}\left[ \nu -w(0)/2 + \varphi\right]$, with $w(1,2)$
playing the role of the bare  propagator and the non local term  $W_{\textrm{(HS)}}$ that of the interaction. 
However purists will note that this is not a ``canonical'' field theory \textit{stricto sensu}
since the coupling between  the field $\varphi$ and the external source $\nu$ is non-linear. 
As detailed at length in refs.~\cite{RG-Fluids,Caillol-RG} this point can be circumvented by the change of variables
$\chi=\varphi+ \nu -w(0)/2 $. By doing so one recovers a ``canonical'' field theory involving a linear coupling between 
$\chi$ and $\nu$.  Moreover one can establish a rigorous
mapping between these two field theories. Then all the knowledge amassed in statistical field theory can be injected
in the theory of liquids, notably well established perturbative technics and/or  more recent  non-perturbative approaches.
Recently however we proposed rather  to  consider the action $\mathcal{S}[\varphi]$ as a well-educated one and work directly with it rather
than to try to scholarly adhere  to a canonical action via the mapping $\varphi \leftrightarrow \chi$.
As discussed in reference~\cite{Caillol-RG}, the adoption of this  ``non-canonical'' point of view leads to  
a remarkably simple reconciliation between 
the points of view of Wilson-Polchinski on the one hand and that
of Wetterich in the other hand. Applied to liquids, as will be done here, it yields a new  
version of the smooth HRT theory with some advantages on my previous attempts \cite{RG-Fluids}.

\section{Coarse graining \label{CO}}
The coarse graining procedure involved to build families of actions with the same physics at long distances is the core
of the RG theory. In this section we introduce this procedure step-by -step  following recent developments 
in field theory \cite{Morris,Caillol-RG} 
\subsection{Ultra-violet regularization of the pair potential\label{UV}}
I noted in section~\ref{GC} that the pair potential (propagator)  $w(r)$ can be regularized at short distances $0<r<\sigma$,
\textit{i. e.} in the ultra-violet (UV) regime, without changing the GCPF $\Xi [ \nu ]$ of the system.
I introduce a potential $w^{\Lambda}_{0}(r)$ defined in Fourier space by
\begin{equation}
 \widetilde{w}^{\Lambda}_{0}(q) = C(q,\Lambda) \widetilde{w}(q) \; ,
\end{equation}
where $ C(q,\Lambda)$ is an UV cut-off which is equal to 1 for $q< \Lambda$ and to 0 for $q> \Lambda$.
Typically $C(q,\Lambda) = 1 - \Theta_{\epsilon}(q,\Lambda)$
where $\Theta_{\epsilon}(q,\Lambda)$ is a function which looks like the step  function $\Theta(q-\Lambda)$. It could
be precisely this function and we would then speak of a sharp cut-off; or it could be a smooth version with $\Theta_{\epsilon}(q,\Lambda)$
varying smoothly from 0 to 1 in a small interval $(\Lambda - \epsilon,\Lambda + \epsilon) $ about the UV cut-off; in this case we would 
speak of a smooth UV cut-off. It is sometimes convenient to choose  $C(q,\Lambda)=\overline{C}(x=q/\Lambda)$ but by
no means compulsory.

In any case if $ \Lambda \sigma \gg 1 $ then  $w(r)$ and  $w^{\Lambda}_{0}(r)$  coincide outside the core and moreover we have
$\widetilde{w}^{\Lambda}_{0}(0)=\widetilde{w}(0)$ and $ \vert w^{\Lambda}_{0}(r=0) \vert < \infty$. 
Although this  UV regularization does not change the GCPF $\Xi\left[ \nu \right] $ we shall  henceforth note this functional
 $\Xi^{\Lambda}_{0}\left[ \nu \right] $ to emphasize that it depends functionally on $w^{\Lambda}_{0}$.
For further  convenience,  we also need a special notation for
the inverse of the  regularized propagator $w^{\Lambda}_{0}$  which will be christened  $R^{\Lambda}_{0} =[w^{\Lambda}_{0}]^{-1}$.
We can thus rewrite~\eqref{csi}  and~\eqref{attractive} as 

\begin{subequations}
\begin{eqnarray}
\label{csi_bis}
\Xi^{\Lambda}_{0}\left[ \nu \right]  &=&
\mathrm{Tr}\left[ \; \exp\left(  \frac{1}{2} \widehat{\rho} \cdot    \widetilde{w}^{\Lambda}_{0}   \cdot \widehat{\rho} 
+\widehat{\rho} \cdot  [ \nu -  w^{\Lambda}_{0} (0)/2  ]
\right) \right] \; ,       \label{blo}\\
 &=&\frac{1}{\mathcal{N}_{w^{\Lambda}_0}}\int \mathcal{D} \varphi \;
\exp \left( -\frac{1}{2}
 \varphi \cdot  R^{\Lambda}_{0} \cdot  \varphi 
 + W_{\textrm{(HS)}}\left[ \nu -w^{\Lambda}_{0} (0)/2 + \varphi\right]
 \right) \; . \label{blu}
\end{eqnarray}
\end{subequations}
Of course $\Xi^{\Lambda}_{0}\left[ \nu \right]  \equiv  \Xi\left[ \nu \right] $ as soon as $  \Lambda \sigma \gg 1 $ which
will be assumed.

\subsection{Two lemma\label{lemmz}}
I prefer, for the convenience of the reader,  to give here two lemma necessary for subsequent developments
rather than to postpone them in an appendix. 
 Let $\Delta$ some definite positive operator and 
$\mathcal{F}[\varphi]$ an arbitrary functional of real scalar field $\varphi$,   then
\begin{equation}
\label{T1}
 \frac{1}{\mathcal{N}_{\Delta}}  \int \mathcal{D} \varphi \;
\exp ( -\dfrac{1}{2} \varphi \cdot \Delta^{-1} \cdot \varphi ) \; 
\mathcal{F}[\varphi + \varphi_0] = \exp(D_0) \mathcal{F}( \varphi_0] \; ,
\end{equation}
where the operator $D_0$ reads
\begin{equation}
 D_0 \ldots = \frac{1}{2} \int_{\Omega} \, d^dx \, d^dy \; \Delta(x,y) \dfrac{\delta^2 \ldots }{\delta \varphi_0(x) \delta \varphi_0(y)} \; .
\end{equation}
An interesting application of lemma~\eqref{T1} is to reconsider equation~\eqref{blu}
as
\begin{equation}
\label{R1}
 \Xi^{\Lambda}_{0}[ \nu +  w^{\Lambda}_{0}(0)/2  ] = e^{D^{\Lambda}_{0}} \; \Xi_{\textrm{(HS)}}\left[ \nu \right] \; ,
\end{equation}
therefore the operator $e^{D^{\Lambda}_{0}}$ constructs the full GCPF  of  the system from the 
GCPF of the reference HS system.

The second lemma is sometimes referred to as Bogolioubov theorem :

 Let $\Delta_i$, $1 \leq i \leq n$,    be n   definite positive operators and 
$\mathcal{F}[\varphi]$ an arbitrary functional of real scalar field $\varphi$,   then
\begin{eqnarray}
\label{T2}
\int \, \prod_{i=1}^{n} \left\{  \dfrac{\mathcal{D}\varphi_i}{\mathcal{N}_{\Delta_i}}  \exp ( -\dfrac{1}{2} \varphi_i \cdot \Delta_i^{-1} \cdot \varphi_i )\right\} 
 \mathcal{F}( \sum_{i=1}^{n} \varphi_i) &=& \int \, \dfrac{\mathcal{D}\varphi}{\mathcal{N}_{\Delta}} \exp ( -\dfrac{1}{2} 
\varphi \cdot \Delta^{-1} \cdot \varphi ) \times \nonumber \\
& \times & 
 \, \mathcal{F}(  \varphi) \; ,
\end{eqnarray}
where $\Delta = \sum_{i=1}^{n} \Delta_i$.

The two lemma~\eqref{T1} and~\eqref{T2} are easy consequences of Wick's theorem.
\subsection{Blocking the action \label{block}}

I now apply the exact RG approach of Tim Morris \cite{Morris,Caillol-RG}  to our non-canonical field theory.
As a consequence of  Bogolioubov  theorem~\eqref{T2} the GCPF
$\Xi^{\Lambda}_0 \left[ \nu \right]$ can be rewritten in terms of two propagators and two fields as
\begin{subequations}\label{Bogo}
\begin{align}
\Xi^{\Lambda}_0\left[ \nu \right] &=  \frac{1}{\mathcal{N}_{w^{k}_{0}} } \, 
 \int \! \mathcal{D}\varphi_{<} \, \exp\left(
-\dfrac{1}{2}\varphi_{<} \cdot R^{k }_{0} \cdot  \varphi_{<}  
\right) \, \Xi^{\Lambda}_{k}\left[\nu - w_0^k(0)/2, \varphi_{<}\right]  \, ,   \label{Mi}\\
\Xi^{\Lambda}_{k}\left[\nu, \varphi_{<} \right] &= \frac{1}{\mathcal{N}_{w^{\Lambda}_{k}} } \, 
 \int \! \mathcal{D}\varphi_{>} \, \exp\left (
-\dfrac{1}{2}\varphi_{>} \cdot R^{\Lambda}_{k} \cdot  \varphi_{>} \right. \; + \nonumber \\
&+ \left.W_{\textrm{(HS)}}\left[\nu  -\dfrac{w^{\Lambda}_k(0)}{2} +\varphi_{<} + \varphi_{>} \right] \right) \, ,  \label{Mo}
\end{align}
\end{subequations}
where $0\leq k\leq \Lambda$ is the running scale of the RG and where
\begin{equation}\label{decompo}
\varphi = \varphi_{<} + \varphi_{>} \textrm{  and  }  w^{\Lambda}_{0}=w^{\Lambda}_{k}+w^{k}_{0} \, .
\end{equation}
In \eqref{Bogo}-\eqref{decompo} I have separated the field $\varphi$ into ``rapid'' ($\varphi_{>}$)
and ``slow'' modes \mbox{($\varphi_{<}$)}. The low-energy modes are associated to the propagator
$w^{k}_{0} $ (with inverse $R^{k}_{0} $) which is cut off from above by $k$, while
 the  high-energy modes are associated to
the propagator $w^{\Lambda}_{k} $         (with inverse $R^{\Lambda}_{k} $) 
which is cut off from below by $k$ and from above by $\Lambda$.
I demand  that 
$\widetilde{w}^{\Lambda}_{k}(q)=\widetilde{w}_{0}(q) (C(q,\Lambda) -C(q,k) ) $ should be positive which
will be assumed henceforth.
In the popular case where $C(q,\Lambda)=\overline{C}(q/\Lambda)$ it is sufficient for the cut-off function $\overline{C}(x)$
to be a monotonous decreasing function of its argument.
Some comments are in order.
\begin{enumerate}
 \item[(i)] In order to establish equations~\eqref{Bogo} I used $ w^{\Lambda}_{0}(0) =w^{\Lambda}_{k}(0)+w^{k}_{0}(0)$
as implied by~\eqref{decompo}.
 \item[(ii)] Note that the ``slow'' and ``rapid'' modes, respec. $\varphi_<$ and $\varphi_>$ are ordinary scalar fields, in
particular they have a full spectrum
of Fourier components $\widetilde{\varphi}_{< }(q) $ and  $\widetilde{\varphi}_{>}(q) $  (even the condition  $0\leq q \leq \Lambda$
is not compulsory if the bare propagator is UV regularized).
The cut-off at scale ``k''  acts only on  the propagators.
 \item[(iii)] The functional $\Xi^{\Lambda}_{k}[\nu, \varphi ] $ noted in this way by Morris \cite{Morris} is the crux of the whole matter
since it allows, as I will show,
 to make explicit the link between  the Wilsonian action and the effective average action.
However  here this link  proves trivial since, as apparent in formula~\eqref{Mo},   $\Xi^{\Lambda}_{k}[\nu, \varphi ] $,
depends functionally  only on the sole variable $\nu + \varphi$.
\end{enumerate}

Let  me first set $\varphi_{<}=0$ in \eqref{Mo}. It yields
\begin{subequations}\label{niceZ}
\begin{align}
\Xi^{\Lambda}_{k}\left[ \nu, \varphi_{<}=0 \right] & \triangleq \Xi^{\Lambda}_{k}\left[ \nu \right]  (  \triangleq
    \exp    \left( W^{\Lambda}_{k}\left[ \nu \right]\right)   )   \label{Wett-a} \\
&= \frac{1}{\mathcal{N}_{w^{\Lambda}_{k}} } \, 
 \int \! \mathcal{D}\varphi \,
 \exp\left(
-\dfrac{1}{2}\varphi \cdot R^{\Lambda}_{k} \cdot  \varphi  \;
+ W_{\textrm{(HS)}}\left[\nu -  w^{\Lambda}_k(0)/2 + \varphi  \right] 
\right) \,\label{Wett-b} \, .
\end{align}
\end{subequations}

From the point of view of field theory this shows  shows that $W^{\Lambda}_{k}\left[\nu  \right] = \ln  \Xi^{\Lambda}_{k}\left[ \nu \right]$ is the 
Helmholtz free energy of the rapid modes
$\varphi_{>} $ in the presence of the source $\nu(x)$; $W^{\Lambda}_{k}$ is thus the generating functional of
connected correlation functions with UV regularization (at $\Lambda$) and an  infra-red (IR) cut-off (\textit{i. e.} at  k).
This functional is related by a Legendre transformation to the effective average action of Wetterich,
 as it willl be discussed  in section~\ref{Flow}.

From the point of view of the theory of liquids  clearly, (see \textit{e. g.} equation~\eqref{attractive}) $ \Xi^{\Lambda}_{k}\left[ \nu \right]$ 
is precisely  the GCPF of
a system of hard spheres with additional pairwise potentials $ w^{\Lambda}_{k}\left[r \right]$, \textit{i. e. },  the k-system,  to adopt the terminology
of Reatto `et al.' \cite{Parola1}. Therefore  one also has :
\begin{equation}
\label{zub}
  \Xi^{\Lambda}_{k}\left[ \nu \right] = \mathrm{Tr}\left[ \; \exp\left(  \frac{1}{2} \widehat{\rho} \cdot    \widetilde{w}^{\Lambda}_{k}   \cdot \widehat{\rho} 
+\widehat{\rho} \cdot  [ \nu -  w^{\Lambda}_{k} (0)/2  ]
\right) \right] \; .
\end{equation}
Note that equation~\eqref{zub} is valid provided  the propagator $\widetilde{w}^{\Lambda}_{k}(x,y)$
(pair potential)  is definite positive (attractive) so that a Hubbard-Stratonovich transform is licit; that is why I imposed
earlier that  the cut-off function  $x \rightarrow \overline{C}(x)$ should be a decreasing function.
Comparing equations~\eqref{Mo} and~\eqref{Wett-b}  one  thus has synthetically
\begin{equation}
 \label{synt}
\Xi^{\Lambda}_{k}\left[ \nu, \varphi \right] = \Xi^{\Lambda}_{k}\left[ \nu + \varphi \right] \; .
\end{equation}

I now introduce the Wilsonian action
\begin{equation}
 \mathcal{S}^{\Lambda}_{k}[\varphi ] \triangleq -  W^{\Lambda}_{k}\left[  \varphi \right] \; ,
\end{equation}
which allows me to rewrite~\eqref{Mi} as 
\begin{equation}
 \label{Wilson}
\Xi^{\Lambda}_0\left[ \nu + w_0^k(0)/2 \right] =  \frac{1}{\mathcal{N}_{w^{k}_{0}} } \, 
 \int \! \mathcal{D}\varphi_{<} \, \exp\left(
-\dfrac{1}{2}\varphi_{<} \cdot R^{k }_{0} \cdot  \varphi_{<}  
- \mathcal{S}^{\Lambda}_{k}[\nu + \varphi_{<}]  \; .
\right) 
\end{equation}
Indeed, in equation~\eqref{Wilson} 
$S^{\Lambda}_{k}\left[\varphi_{<} \right]$   play the role of the effective action of the 
slow modes at scale $k$ \cite{Wilson,Wegner,Morris}.  Here $k$ plays the role of an UV cut-off.
Note that the functional identity $S^{\Lambda}_{k} = - W^{\Lambda}_{k}$ is
not true for a canonical field theory \cite{Morris}.

We end this section by applying lemma~\eqref{T1} to the 2 equations~\eqref{Bogo} which can thus be rewritten as
\begin{subequations}
 \begin{eqnarray}
  \Xi^{\Lambda}_{0}[ \nu + w^{k}_{0}(0)/2 ] &=&e^{D^{k}_{0}} \; \Xi^{\Lambda}_{k}[\nu] \;  \\
 \Xi^{\Lambda}_{k}[ \nu + w^{\Lambda}_{k}(0)/2 ] &=&e^{D^{\Lambda}_{k}} \; \Xi_{\textrm{(HS)}}[\nu]  \label{solu}\; 
 \end{eqnarray}
\end{subequations}
from which it follows that
\begin{equation}
   \Xi^{\Lambda}_{0}[ \nu + w^{k}_{0}(0)/2 ]= e^{D^{k}_{0}}  e^{D^{\Lambda}_{k}} \Xi_{\textrm{(HS)}}[\nu] \; .
\end{equation}
Comparing with equation~\eqref{R1} yields
\begin{equation}
   e^{D^{\Lambda}_{0}}    =    e^{D^{k}_{0}}  e^{D^{\Lambda}_{k}} \; .
\end{equation}
This is the semi-group law of the RG.
\section{RG Flow equations \label{Flow}}
\subsection{RG flow of the grand potential  $W^{\Lambda}_{k}$\label{flowW}}
From the definitions~\eqref{decompo}  when $k \to \Lambda$  then $ w^{\Lambda}_{k}(r) \to 0$ and the k-system 
at $k=\Lambda$ is a fluid of hard spheres. When $k$ decreases from $\Lambda$ to $k=0$
 more and more Fourier components are included in the potential $w^{\Lambda}_{k}(r) $ and finally
precisely at $k=0$ the k-system coincides with the full model since $w^{\Lambda}_{k}(r) $ tends
to $w^{\Lambda}_{0}(r) $, \textit{i. e. } essentially $w(r)$ since both potentials differ only in the hard cores
with no effect on the physics.
In this section I establish the equations which govern the flow of the grand potential  $W^{\Lambda}_{k}$
introduced in Sec~\ref{CO} and its Legendre transform. The flow equations of the Wilsonian action 
 $\mathcal{S}^{\Lambda}_{k}$ follows trivially.
Note that  the formal solution of the not yet established flow equation for $W^{\Lambda}_{k}$ is already known,
it is given by equation~\eqref{solu}.

The  flow equation for $W^{\Lambda}_{k}$  can be  obtained in the framework of statistical field theory 
and, apart some tricks due to the self-energies, along  essentially the same lines as in  reference~\cite{Caillol-RG}. 
I  postpone this derivation  to  appendix~\ref{appendixA} and I give here only  the proof in the framework of liquid theory.
An easy task. The starting point is equation~\eqref{zub} which I differentiate with respect to k at fixed $\nu(x)$. It gives  readily
\begin{equation}
 \label{flowW1}
 \left. \partial_k W^{\Lambda}_{k}[\nu] \right\arrowvert_{\nu}= \frac{1}{2} \int_{\Omega}d^dx \, d^dy \,
       \partial_k  w^{\Lambda}_{k} (x,y)         
      \left\{  G^{\Lambda}_{k}(x,y) -   \rho^{\Lambda}_{k}(x)\delta^{(d)}(x-y)   \right\}
\end{equation}
where  $ \rho^{\Lambda}_{k}(x)=< \widehat{\rho}(x|\mathcal{C})>_{\textrm{GC}}$  denotes the mean density of the k-system
in the GC ensemble and  $G^{\Lambda}_{k}(x,y)= <\widehat{\rho}(x|\mathcal{C}) \;  \widehat{\rho}(y|\mathcal{C}) >_{\textrm{GC}}$ 
its pair correlation function at chemical potential $\nu(x)$.  Since  one has
$  \rho^{\Lambda}_{k}(x)= \delta W^{\Lambda}_{k}/\delta \nu(x) \triangleq W^{(1) \; \Lambda}_{k}(x) $ and  also 
$G^{(T) \; \Lambda}_{k}(x,y) =G^{(T) \; \Lambda}_{k}(x,y) - \rho^{\Lambda}_{k}(x) \rho^{\Lambda}_{k}(y)=
\delta^{(2)} W^{\Lambda}_{k}/\delta \nu(x) \delta \nu(y)  \triangleq W^{(2) \; \Lambda}_{k}(x,y)  $
 one can rewrite equation~\eqref{flowW1} under the closed form
\begin{eqnarray}
 \label{flowW2}
 \left. \partial_k W^{\Lambda}_{k}[\nu +w^{\Lambda}_{k} (0)/2  ] \right\arrowvert_{\nu}&=& \frac{1}{2} \int_{\Omega}d^dx \, d^dy \,
       \partial_k  w^{\Lambda}_{k} (x,y)    
      \{  W^{(2) \; \Lambda}_{k}(x,y) \nonumber \\
 &+& W^{(1) \; \Lambda}_{k}(x) W^{(1) \; \Lambda}_{k}(y) \}  \; .
\end{eqnarray}

This partial differential equation (PDE) must be supplemented by the initial condition
 $W^{\Lambda}_{k} \equiv W_{\textrm{(HS)}}$.
This equation is closed only superficially since the pair correlation function $W^{(1) \; \Lambda}_{k}(x)$
and  $W^{(2) \; \Lambda}_{k}(x,y)$ depends functionally on the chemical potential $\nu(x)$. 
Differentiating  functionally successively both members
of equation~\eqref{flowW2} with respect to the field $\nu(x)$ one obtains a hierarchy, or tower, of
equations for the $W^{(n) \; \Lambda}_{k}(x_1, \ldots, x_n) =
 \delta^n  W^{ \Lambda}_{k}/\delta \nu(x_1) \ldots \delta \nu(x_n)$,
\text{i. e. } the Green's -or connected correlation-  functions. 

\subsection{RG flow of the Wilsonian action  $S^{\Lambda}_{k}$\label{flowSS}}
The flow equation~\eqref{flowW2} has the structure of Wilson-Polchinski's equation for the Wilsonian action;
 indeed with $S^{\Lambda}_{k} = - W^{\Lambda}_{k}$ it takes the usual form \cite{Polchinski,Bervillier,Morris}
\begin{eqnarray}
 \label{flowS}
 \left. \partial_k S^{\Lambda}_{k}[\varphi + w^{\Lambda}_{k} (0)/2  ] \right\arrowvert_{\varphi}&=& \frac{1}{2} \int_{\Omega}d^dx \, d^dy \,
       \partial_k  w^{\Lambda}_{k} (x,y)         
      \{  S^{(2) \; \Lambda}_{k}(x,y)     \nonumber \\
&-& S^{(1) \; \Lambda}_{k}(x) S^{(1) \; \Lambda}_{k}(y) \}.
\end{eqnarray}
where $S^{(n) \; \Lambda}_{k}(x_1, \ldots ,x_n) = \delta^{n}S^{ \Lambda}_{k}/\delta \varphi(x_1) \ldots \delta \varphi(x_n)$.
This equation, which  must be supplemented with the initial condition $S^{ \Lambda}_{\Lambda}= -W_{\textrm{(HS)}}$, does not appear 
to have yet been considered in the theory of liquids contrary to the abundant literature devoted to it 
in statistical field theory (see \textit{e. g.} \cite{Bervillier} and references quoted herein). 

\subsection{RG flow of the effective average Kohn-Sham free energy  $\beta \mathcal{A}^{\Lambda}_{k}$\label{flowAA}}

Since the seminal works of Reatto `et al.' \cite{paro1, paro2,Parola,Parola1}
 it is usual to consider rather the flow for the Kohn-Sham free energy 
$\overline{A}^{\Lambda}_{k}[\rho]$ of the k-system which is defined as the Legendre
transform of $W^{\Lambda}_{k}[\nu] $.
 One has the usual couple of relations
\begin{subequations}
\begin{eqnarray}
 W^{\Lambda}_k \left[\nu \right]&=& \sup_{\rho}\left\lbrace 
 \nu \cdot \rho  - \beta \overline{\mathcal{A}}^{\Lambda}_k\left[ \rho \right] \right\rbrace \; \;  \forall \nu \; , \label{P1} \\
\beta \overline{\mathcal{A}}^{\Lambda}_k \left[ \rho\right] &=& \sup_{\nu }\left\lbrace 
 \rho \cdot  \nu  - W^{\Lambda}_k\left[\nu \right]  \right\rbrace \; \; \forall \rho  \; .\label{P2} 
\end{eqnarray}
\end{subequations}
Obviously the flow equation for  $\beta \overline{\mathcal{A}}^{\Lambda}_k\left[ \rho \right]$ should be deduced
from that of  $ W^{\Lambda}_k \left[\nu \right]$. 
Let me consider for instance equation~\eqref{P1}. I have for all $\nu$
\begin{equation}
\label{ul}
  W^{\Lambda}_k \left[\nu \right] = \rho^{\star}\cdot \nu - \beta \overline{\mathcal{A}}^{\Lambda}_k\left[  \rho^{\star} \right] \; ,
\end{equation}
where $ \rho^{\star}(x)$ is, if it does exists,  the unique  solution of the implicit stationnary equation 
\begin{equation}
\label{statio}
 \nu = \left. \dfrac{\delta \beta \overline{\mathcal{A}}^{\Lambda}_k \left[ \rho\right] }{\delta \rho(x)}\right|_{\rho =\rho^{\star}}
\end{equation}
Differentiating~\eqref{ul} at fixed $\nu$ gives 
\begin{equation}
\left. \partial_k   W^{\Lambda}_k \left[\nu \right]\right|_{\nu} = \partial_k  \rho^{\star} \cdot \nu -
\left.\dfrac{\delta \beta \overline{\mathcal{A}}^{\Lambda}_k\left[  \rho^{\star} \right] }{\delta \rho^{\star}}\right|_{k} \cdot
\partial_k \rho^{\star} - \left. \partial_k  \beta \overline{\mathcal{A}}^{\Lambda}_k\left[  \rho^{\star} \right] \right|_{\rho^{\star}} \; ,
\end{equation}
which further simplifies thanks to the stationnary condition~\eqref{statio} with the final result :
\begin{eqnarray}
\label{hh}
  \left. \partial_k \overline{\mathcal{A}}^{\Lambda}_{k}[\rho^{\star}] \right\arrowvert_{\rho^{\star}}
&=& -  \left. \partial_k W^{\Lambda}_{k}[\nu] \right\arrowvert_{\nu} \; .
\end{eqnarray}
 Similarly, starting instead  from equation~\eqref{P2} one finds that 
\begin{eqnarray}
\label{hhh}
  \left. \partial_k \overline{\mathcal{A}}^{\Lambda}_{k}[\rho] \right\arrowvert_{\rho}&=& - 
 \left. \partial_k W^{\Lambda}_{k}[\nu^{\star}] \right\arrowvert_{\nu^{\star}} \; , 
\end{eqnarray}
where $\nu^{\star}$ is the unique chemical potential which is, if it exists, solution of equation~\eqref{P2}. I shall drop the subscript ''$\star$'' in the sequel.

At this point I  introduce the direct correlation functions
\begin{equation}
 \label{direct}
 \overline{\mathcal{C}}^{(n) \,\Lambda}_{k}(x_1, \ldots, x_n)=-
\delta^{n}{\overline{\mathcal{A}}^\Lambda}_{k}/\delta \rho(x_1) \ldots \delta \rho(x_n) \; ,
\end{equation}
which are the analogous to  the vertex functions in field theory. I have simplified the notations and it must be stressed
that, in addition to their arguments these function depend functionally on the profile $\rho(x)$.
The  $\overline{\mathcal{C}}^{(n) \,\Lambda}_{k}(x_1, \ldots, x_n)$
and the  $W^{(n) \,\Lambda}_{k}(x_1, \ldots, x_n)$ are linked by generalized Ornstein-Zernike (OZ) equations and in particular 
one has  the usual OZ equation : $W^{(2) \,\Lambda}_{k}(x, y)= - [\overline{\mathcal{C}}^{(2) \,\Lambda}_{k}]^{-1}(x,y)$ \cite{Parola1}.
This last property allows to rewrite equation~\eqref{hh} in a closed form involving $\overline{\mathcal{A}}^{\Lambda}_{k}$,
its functional derivatives and the propagator. One finds :
\begin{align}
\label{bof}
 \left. \partial_k \overline{\mathcal{A}}^{\Lambda}_{k}[\rho]\right|_{\rho}  = & - \frac{1}{2}\int_{\Omega}\, d^dx \, d^dy \,
                      \partial_k w^{k}_{0}(x,y) [\overline{\mathcal{C}}^{(2) \,\Lambda}_{k}]^{-1}(x,y)  \nonumber \\
  & + \frac{1}{2} \int_{\Omega}\, d^dx \, d^dy \, \partial_k w^{k}_{0}(x,y) \rho(x) \rho(y) 
     - \frac{1}{2} \int_{\Omega}\, d^dx \,  \partial_k w^{k}_{0}(x,x) \rho(x) \; .
\end{align}
Note that, in order to obtain the result I  made use of the trick $\partial_k w^{k}_{0}(x,y)=-\partial_k w^{\Lambda}_{k}(x,y)$.
However, the resulting equation~\eqref{bof}  is quite awkward and it is convenient,
by adapting  the ideas of  Reatto `et al.' \cite{paro1, paro2,Parola,Parola1} to our case, to introduce a modified
Kohn-Sham free energy :
\begin{equation}
\label{def}
\beta \mathcal{A}^{\Lambda}_k \left[ \rho\right]=\beta \overline{\mathcal{A}}^{\Lambda}_k \left[ \rho\right]
-\frac{1}{2} \rho \cdot w^{k}_{0} \cdot \rho +\frac{1}{2} \rho \cdot w^{k}_{0}(0) \; .
\end{equation}
This new functional was introduced by Wetterich `et al.' \cite{Wetterich} in statistical field theory
under the name of ``the effective
average action'' independently from Reatto  `et al.'; it differs from $\beta \overline{\mathcal{A}}^{\Lambda}_k \left[ \rho\right]$
by a simple quadratic form and satisfies obviously the following flow equation 
\begin{subequations}
\begin{eqnarray}
\label{flowA}
 \partial_k \mathcal{A}^{\Lambda}_{k}[\rho] &=&- \frac{1}{2}  \int_{\Omega}d^dx \, d^dy \,
\partial_{k} w^{k}_{0}(x,y) \; \left\{
\mathcal{C}^{(2) \, \Lambda}_{k}-  w^{k}_{0}
 \right\}^{-1}(x,y) \\
\mathcal{C}^{(n) \, \Lambda}_{k}(x_1,\ldots, x_n) &=& - \dfrac{\delta^{n}  \mathcal{A}^{\Lambda}_{k}[\rho] }
{\delta \rho(x_1) \ldots \delta \rho(x_n)  }
\end{eqnarray}
\end{subequations}
I already obtained this equation in my  first version of the RG group for liquids
but with the help of a canonical field theory \cite{RG-Fluids}.
Note that the direct correlation functions $\mathcal{C}^{(n) \, \Lambda}_{k}(x_1,\ldots, x_n)$
differ from the  $ \overline{\mathcal{C}}^{(n) \,\Lambda}_{k}(x_1, \ldots, x_n)$ only for $n \geq 3$; in
particular one has 
\begin{equation}
\label{pos}
 \mathcal{C}^{(2) \, \Lambda}_{k}(x,y) = \overline{\mathcal{C}}^{(2) \, \Lambda}_{k}(x,y) + w^{k}_{0}(x,y) \; .
\end{equation}
I stress that although $\beta \overline{\mathcal{A}}^{\Lambda}_k \left[ \rho\right]$ is a convex functional 
of the profile $\rho(x)$ this is not the case in general for $\beta \mathcal{A}^{\Lambda}_k \left[ \rho\right]$
except of course at $k=0$ since
 $\beta \mathcal{A}^{\Lambda}_{k=0}[\rho] = \beta \overline{\mathcal{A}}^{\Lambda}_{k=0}\left[ \rho\right]$ as 
apparent in formula~\eqref{def}. Indeed although  $\overline{\mathcal{C}}^{(2) \, \Lambda}_{k} <0$ in the sense of operators,
equation~\eqref{pos} shows that  $ \mathcal{C}^{(2) \, \Lambda}_{k}$ could become positive due to the addition of the
positive operator $w_0^k$.

 Equation~\eqref{flowA} must be supplemented with an initial condition at $k=\Lambda$. From 
$W^{\Lambda}_{\Lambda}= W_{\textrm{(HS)}}$ as follows from equation~\eqref{solu} and $D^{\Lambda}_{\Lambda}=0$
one concludes  that $\beta \overline{\mathcal{A}}^{\Lambda}_{\Lambda}=\beta \mathcal{A}_{\textrm{(HS)}}$
and thus from~\eqref{def}
\begin{equation}
\beta \mathcal{A}^{\Lambda}_{\Lambda}[\rho]= \beta \mathcal{A}_{\textrm{(HS)}} [\rho] 
 -\frac{1}{2} \rho \cdot w^{\Lambda}_{0} \cdot \rho + \frac{1}{2} \rho \cdot w^{\Lambda}_{0}(0) \; .
\end{equation}
In KSSHE theory the initial conditions are identical but  are obtained in a more complicated way which requires
the divergence of the regulator at $k=\Lambda$ \cite{RG-Fluids}.

It turns out that the initial Kohn-Sham free  energy of the k-system, \textit{i. e. } at $k=\Lambda$,
coincides with the mean-field Kohn-Sham free 
energy at \textit{any} scale ``k'', \textit{i. e.} one has 
$ \beta \overline{\mathcal{A}}_{k}^{\Lambda \, (\textrm{MF} )}[\rho]= \beta \mathcal{A}^{\Lambda}_{\Lambda}[\rho]$
for all k.
Moreover this MF free energy  constitutes a rigorous upper bound to
the exact free energy and thus              
$ \beta \mathcal{A}_{\Lambda}^{\Lambda }[\rho] \geq \beta \mathcal{A}_{k}^{\Lambda}[\rho]$.
This result  already obtained  in references~\cite{KSSHE,RG-Fluids} for the KSSHE 
theory is proved for this new version of the theory  in appendix~\ref{appendixB} where a precise definition of the MF
approximation is provided.

For a homogeneous system $\beta \mathcal{A}^{\Lambda}_{k}\left[ \rho \right]=V f^{\Lambda}_k(\rho)$
where $\beta$ times the free energy per unit volume $f^{\Lambda}_k(\rho)$ is a function (not a functional) of $\rho$ and the flow equation for
 $f^{\Lambda}_k(\rho)$ is easily deduced from equation~\eqref{flowA} and reads 
\begin{eqnarray}
\label{flow-f}
\partial_{k}f^{\Lambda}_k &=&
-  \frac{1}{2} \int_{q} \frac{\partial_{k} \widetilde{w}^{k}_{0}(q)
}{\widetilde{C}_{k}^{(2) \, \Lambda } (q) - \widetilde{w}^k_0} \; ,
\end{eqnarray}
where $\int_q \equiv \int d^{d} q/(2 \pi)^d$ and $\widetilde{C}^{(2) \, \Lambda}_{k} (q)$ is the Fourier transform of 
$C^{(2) \, \Lambda}_k(c,y) \equiv C^{(2) \, \Lambda}_k(r= \Arrowvert x-y \Arrowvert)$ for a translationally invariant fluid.
Note that the right hand side is non singular since the denominator is negative definite.
It is a general property that the flow of $f_k^{\Lambda}$ has neither UV or IR singularities, in particular IR singularities (near a critical point) are
smoothened by k and they build up progressively as the scale-k is lowered\cite{Wetterich,Delamotte}.
.

To make contact with the work of Reatto `et al.' let me consider now 
 the case of a sharp cut-off, \textit{i. e. } $C(q,k) = 1 - \Theta(q-k)$. Then 
equation~\eqref{flow-f} becomes \cite{Caillol-RG}.
\begin{eqnarray}
\label{flow-sharp}
\partial_{k} f^{\Lambda}_k &=&
 \frac{k^{d-1}}{2}  \frac{S_d}{(2 \pi)^d} \ln  \{  1
- \frac{\widetilde{w}^{\Lambda}_0(k)}{ \widetilde{C}^{(2) \, \Lambda}_{k} (k)}\} \; ,
\end{eqnarray}
where $S_d = 2 \pi^{d/2}/\Gamma(d/2)$. This  is the usual  HRT equation \cite{Parola1}.  
 In order to obtain \eqref{flow-sharp} from 
\eqref{flow-f} one replaces the sharp $\Theta(q-k)$ by a smooth  one, $\Theta_{\epsilon}(q,k)$,
denotes  $\delta_{\epsilon}(q,k) = -\partial_k \Theta_{\epsilon}(q,k) $ the smooth $\delta$-function
and finally  make use of Morris lemma which states that, for $\epsilon \to 0$ 
\begin{equation}
\label{lemme}
\delta_{\epsilon}(q,k)f( \Theta_{\epsilon}(q,k),k) \to
\delta(q-k) \; \int_{0}^{1}dt \; f(t,q) \;,
\end{equation}
provided that the function $f( \Theta_{\epsilon}(q,k),k)$ is continuous at $k=q$ in the limit $\epsilon \to 0$,
which is the case here. In the canonical field KSSHE theory for liquids one needs to consider rather
an ultra-sharp cut-off regulator \cite{Caillol-RG}.

Whatever the considered version of  HRT, smooth or sharp cut-off, \textit{i. e.} respec.
equations~\eqref{flow-f} and \eqref{flow-sharp}, both equations are of 
a formidable complexity despite their apparent simplicity.
 Indeed  the kernels of these PDE involve the two-body 
 $C^{(2)\, \Lambda}_{k}\left[\rho \right]$ which depends \textbf{functionally}
on the density. Clearly one can deduce from these equations an infinite tower of equations for
 the $C^{(n)\, \Lambda}_{k}\left[\rho \right](1, \ldots,n)$ by differentiating them functionnaly with respect
to the density profile $\rho(x)$. A discussion of this point is however out the scope of the present paper,
it has been developed in detail in a general context in reference \cite{Caillol-RG}.

\section{Conclusion \label{conclu}}

In this paper I have developed a new version of the ``smooth'' hierarchical reference theory
for liquids. It can be seen as an effort to conciliate the point of view of the theory
of liquids and that of statistical field theory. Technically it is an application to a peculiar
field theory (KSSHE), aimed at representing liquids, of the more general theory developed in
reference~\cite{Caillol-RG}. The equivalence pair potentials $ \leftrightarrow$ propagators is the crux
of the whole matter.  In addition, the interplay between the points of view of Wilson and Wetterich
in  their  application to the theory of liquids yields a better understanding of the work of Reatto and
his collaborators.

From a prosaic point of view the  differences
with the first version of this work  \cite{RG-Fluids} are small but with interesting consequences. In practice,
if one leaves  aside philosophical considerations on Wilsonian and Wetterich actions, the main changes
concern essentially the expression of  the pair potential $w^{\Lambda}_{k}$ of the k-system. 
In the first version, following Wetterich, I add a mass
term $\varphi \cdot \mathcal{R}_{k} \cdot \varphi /2$ to the KSSHE action. With 
$\mathcal{R}_{k}(q) \sim Z k^2 (1 - \Theta_{\epsilon}(q,k))$, $Z$ large.
This results in
an ugly  pair potential for the k-system, of the form $\widetilde{w}^{\Lambda}_{k}(q) = \widetilde{w}^{\Lambda}_{0}(q)
/(1 +   \widetilde{w}^{\Lambda}_{0}(q)      \widetilde{\mathcal{R}}_{k}(q) )$. In order to recover Reatto's
equation~\eqref{flow-sharp} one then needs to consider an ultra-sharp cut-off limit, \text{i. e.} : 
$\epsilon \to 0$ and $Z \to  \infty$.  I recall that, here, only the simple sharp cut-off limit $\epsilon \to 0$, ($Z=1$)
is required.

In the present formalism  well-chosen  cut-off functions such as $\overline{C}(x)=\exp(-x^2/2)$ yields
analytical $w^{\Lambda}_{k}(r)$ in many cases, for instance $w(r) = \exp(- a r)/r$ (Yukawa), $ w(r)=1/r^{d+n}$,
etc. Indeed one then obtains  for $w^{\Lambda}_{k}(r)\equiv w_{\textrm{Ewald}}(r)$ nothing but the Ewald potential in direct space,
 the guy of numerical simulations. The lacking contribution in Fourier space corresponds to interactions between ``blocks''
of size $1/k$.
This potential $w_{\textrm{Ewald}}(r)$ is monotonous in general. In the old version the potential $w^{\Lambda}_{k}(r)$ is
 not analytical and oscillates in direct space.

A last difference  can be emphasized : in the old version the initial condition for  the flow equation of
$\beta \mathcal{A}^{\Lambda}_{k}$ requires that the cut-off function 
$\mathcal{R}_{\Lambda} = \infty$, which is difficult to implement rigorously in practice and could lead
to unexpected errors in numerical applications. In the present version, as discussed in section~\ref{flowAA} 
there is no such problem and the initial condition is easily implemented.

Despite the beauty and success of Reatto's HRT some drawbacks of the theory appeal a smooth cut-off version of it,
such as the one exposed here. A sharp cut-off regulator induces singularities in the pair potential  $w^{\Lambda}_{k}(r)$
which decays slowly as  $\sim \cos(k r)/r$ for $r \to \infty$.  This circumstance makes it difficult to solve exactly the integral equations
involved in the most widely used closures of the Hierarchy; one has to resort to approximations which, as discussed by
Reiner \cite{Reiner} are sometimes not easy to controll.
A second drawback of the sharp cut-off version is the value of some critical exponents, notably that of the specific heat,
 $\alpha \sim -0.05$, which is negative, while positive as it should in the smooth cut-off version \cite{Caillol}.

It seems to me that solving exactly the smooth cut-off version of HRT is possible;
that is coupling the numerical solution of a PDE (flow equation) and integral equations (closure of the hierarchy).
Some of the drawbacks of the Reatto's `et al' version of HRT could then be cured.

\section*{Acknowledgments}
This paper was written in honour of  L. Reatto.

\appendix
\label{appendixA}
\section{Alternative derivation of the RG flow  of  $W^{\Lambda}_{k}$}

I derive the flow of $W^{\Lambda}_{k}$ in the framework of field theory (see also reference\cite{Caillol-RG}).
I will start from equation~\eqref{solu}
\begin{eqnarray}
\label{ploc}
 L &=& R \nonumber \\
 \Xi^{\Lambda}_{k}[ \nu_k  = \nu + w^{\Lambda}_{k}(0)/2 ]  &= &e^{D^{\Lambda}_{k}} \; \Xi_{\textrm{(HS)}}[\nu] \; ,
\end{eqnarray}
and take the partial derivatives of the both sides ``$L$'' and  ``$R$''  of  EQ.~\eqref{ploc}
with respect to the  scale ''k''.
One  thus has for the  l.h.s.
\begin{align}
\label{ze1}
 \left. \partial_k L\right|_{\nu} &= \partial_k \, \Xi^{\Lambda}_{k}[\nu_k] +\int_{\Omega} d^dx \, \dfrac{\delta \Xi^{\Lambda}_{k}}
{\delta \nu_k} \partial_k \nu_k \; , \nonumber \\ 
&= \Xi^{\Lambda}_{k} [\nu_k]  \times   \{ \partial_k \, W^{\Lambda}_{k}[\nu_k] +\frac{1}{2}\int_{\Omega}d^dx \,
   W^{(1) \, \Lambda}_{k}[\nu_k] (x) \partial  w^{\Lambda}_{k}(0)/2 \} \; ,
\end{align}
 and, for the  r.h.s. 
\begin{align}
\label{ze2}
\left. \partial_k R\right|_{\nu} &= \partial_k \,D^{\Lambda}_{k} \,e^{D^{\Lambda}_{k}} \Xi_{\textrm{(HS)}}[\nu] \nonumber \\
  &=\frac{1}{2} \int_{\Omega}d^dx \,d^dy \, \partial_k w^{\Lambda}_{k}(x,y) \, \,
\dfrac{\delta^2  \Xi^{\Lambda}_{k}[\nu_k] }{\delta \nu_k(x) \delta \nu_k(y)} \nonumber \\
&=  \Xi^{\Lambda}_{k}[\nu_k] \, \times \frac{1}{2}\int_{\Omega}d^dx \, d^dy \,  \partial_k w^{\Lambda}_{k}(x,y) \,
  G^{\Lambda}_{k}[\nu_k] (x,y)  \; ,
\end{align} 
where I used the expression :
\begin{equation}
D^{\Lambda}_{k} \ldots = \frac{1}{2} \int_{\Omega} \, d^dx \, d^dy \; w^{\Lambda}_{k}(x,y) \dfrac{\delta^2 \ldots }{\delta \nu_k(x) \delta \nu_k(y)} \; .
\end{equation}
Equating equations~\eqref{ze1} and~\eqref{ze2} and noting that the equality, obtained for arbitrary $\nu_k$,  is then valid
in the change $\nu_k \rightarrow \nu$, $\nu$ arbitrary,  one obtains finally
\begin{equation}
 \label{flowW11}
 \left. \partial_k W^{\Lambda}_{k}[\nu] \right\arrowvert_{\nu}= \frac{1}{2} \int_{\Omega}d^dx \, d^dy \,
       \partial_k  w^{\Lambda}_{k} (x,y)         
      \left\{  G^{\Lambda}_{k}(x,y) -   \rho^{\Lambda}_{k}(x)\delta^{(d)}(x-y)   \right\} \; ,
\end{equation}
which indeed coincides with equation\eqref{flowW1}.
\label{appendixB}
\section{Mean field theory at scale ``k'' }
The Mean-Field approximation of the GCPF of  the k-system will be defined as
\begin{equation}
 \Xi_{k}^{\Lambda \, \textrm{(MF)} }[\nu]\triangleq \exp\left\{ -\frac{1}{2} 
\varphi^{\star} \cdot R^{\Lambda}_{k}\cdot  \varphi^{\star} + W_{\textrm{(HS)}}
[ \nu_k +  \varphi^{\star} ]\right\} \; ,
\end{equation}
where $\nu_k = \nu - w^{\Lambda}_{k}(0)/2$ and $ \varphi^{\star}$ is the location of the saddle point 
integrandl~\eqref{Wett-b}. $ \varphi^{\star}$ satisfies the implicit relation 
\begin{eqnarray}
\label{statio_ap}
 R^{\Lambda}_{k} \cdot \varphi^{\star} &=& \dfrac{\delta}{\delta \varphi^{\star} } W_{\textrm{(HS)}}
[ \nu_k +  \varphi^{\star} ] \nonumber \\
&=& \rho_{\textrm{(HS)}}
[ \nu_k +  \varphi^{\star} ](x) \; .
\end{eqnarray}

For a given chemical potential $\nu(x)$ the MF profile of the k-system is given by
$\rho_{k}^{\Lambda \, (\textrm{MF} )}(x)= \delta \ln  \Xi_{k}^{\Lambda \, (\textrm{MF} )}[\nu]
/\delta \nu(x)= \rho_{\textrm{(HS)}}
[ \nu_k +  \varphi^{\star} ](x)$ where I made use of the stationary condition~\eqref{statio_ap}.
Therefore the ``true'' Kohn-Sham free energy of the k-system is given by
 \begin{equation}
  \beta \overline{\mathcal{A}}_{k}^{\Lambda \, (\textrm{MF} )}= - \ln  \Xi_{k}^{\Lambda \, (\textrm{MF} )} + 
\nu \cdot \rho_{k}^{\Lambda \, (\textrm{MF} )} \; .
 \end{equation}
A short calculation will show that \cite{RG-Fluids,Caillol-RG}
\begin{equation}
 \label{MF1}
 \beta \overline{\mathcal{A}}_{k}^{\Lambda \, (\textrm{MF} )}[\rho]=\beta \mathcal{A}_{\textrm{(HS)}}[\rho]
-\frac{1}{2} \rho \cdot w^{\Lambda}_{k} \cdot \rho 
+\frac{1}{2} \rho \cdot w^{\Lambda}_{k}(0) \; . 
\end{equation}
where $ \beta \mathcal{A}_{\textrm{(HS)}}[\rho]$ is the free energy functional of the HS fluid at same 
density. Therefore one finds  for the average effective Kohn-Sham free energy
\begin{equation}
 \label{MF2}
 \beta \mathcal{A}_{k}^{\Lambda \, (\textrm{MF} )}[\rho]=\beta \mathcal{A}_{\textrm{(HS)}}[\rho]
-\frac{1}{2} \rho \cdot w^{\Lambda}_{0} \cdot \rho 
+\frac{1}{2} \rho \cdot w^{\Lambda}_{0}(0) \; . 
\end{equation}
Note that  $\beta \mathcal{A}_{k}^{\Lambda \, (\textrm{MF} )}[\rho]$ is independent
of scale ``k'' and thus equal to its initial value
 $\beta \mathcal{A}_{\Lambda}^{\Lambda \, (\textrm{MF} )}[\rho]$.

I prove now that $\beta \overline{\mathcal{A}}_{k}^{\Lambda \, (\textrm{MF} )}[\rho]$ is a 
rigorous upper bound to the Kohn-Sham free energy of the k-system \cite{Cai-dft,KSSHE}. We consider
the Legendre-Fenchel relation for the reference system of hard spheres :
\begin{equation}
  W_{\textrm{(HS)}} \left[\nu \right]= \sup_{\rho}\left\lbrace 
\nu \cdot \rho  - \beta \mathcal{A}_{\textrm{(HS)}}\left[ \rho \right] \right\rbrace \; \;  (\,\forall \nu\, )
\end{equation}
which implies the Young inequalities 
\begin{equation}
 (\, \forall \nu  \;  ,  \forall \rho \, )  \; \;   W_{\textrm{(HS)}} \left[\nu \right] +  \beta \mathcal{A}_{\textrm{(HS)}}\left[ \rho \right] 
\geq \rho \cdot \nu \;.
\end{equation}
Injecting this inequality in the espression~\eqref{Wett-b} of $ \Xi^{\Lambda}_{k}\left[ \nu \right]$  one gets 
$(\,\forall \nu\, , \forall \rho)$  
\begin{align}
 \Xi^{\Lambda}_{k}\left[ \nu \right] &\geq \exp ( -\beta \mathcal{A}_{\textrm{(HS)}}[\rho] + \rho \cdot \nu_k) \,
\int \!  
\frac{\mathcal{D}\varphi}{\mathcal{N}_{w^{\Lambda}_{k}}} \, \exp(-\dfrac{1}{2}\varphi \cdot R^{\Lambda}_{k} \cdot  \varphi  \; + \rho \cdot \varphi ) \nonumber \\ 
&\geq \exp(  -\beta \mathcal{A}_{\textrm{(HS)}}[\rho] + \rho \cdot \nu_k +\dfrac{1}{2} \rho \cdot w^{\Lambda}_{k} \cdot \rho   ) \; ,
\end{align}
where I made use of Wick's theorem. Rearranging the terms one thus has 
\begin{equation}
 (\,\forall \nu\, , \forall \rho)  \; \;  \rho \cdot \nu - W^{\Lambda}_{k}\left[ \nu \right] \leq 
 \beta \overline{\mathcal{A}}_{k}^{\Lambda \, (\textrm{MF} )}[\rho] \; ,
\end{equation}
and thus 
\begin{equation}
  ( \forall \rho) \; \; \beta \overline{\mathcal{A}}_{k}^{\Lambda \, (\textrm{MF} )}[\rho] \geq \sup_{\nu}
\{   \rho \cdot \nu - W^{\Lambda}_{k}\left[ \nu \right]\} \; .
\end{equation}
Hence, from the very definition of the Legendre transform~\eqref{P2}
\begin{equation}
   ( \forall \rho) \; \; \beta \overline{\mathcal{A}}_{k}^{\Lambda \, (\textrm{MF} )}[\rho] \geq \beta \overline{\mathcal{A}}_{k}^{\Lambda}[\rho]
\end{equation}
Turning now our attention to the effective average Kohn-Sham free energy I find the exact upper bound :
\begin{equation}
   ( \forall \rho) \; \; \beta \mathcal{A}_{k}^{\Lambda \, (\textrm{MF} )}[\rho]=  \beta \mathcal{A}_{\Lambda}^{\Lambda }[\rho] \geq \beta \mathcal{A}_{k}^{\Lambda}[\rho].
\end{equation}


\end{document}